\documentclass{article}

\RequirePackage[OT1]{fontenc}
\RequirePackage{amsthm,amsmath}
\RequirePackage{natbib}
\RequirePackage{graphicx}
\RequirePackage{dsfont}
\RequirePackage{amsmath}
\usepackage{caption}
\usepackage{subcaption}
\usepackage{authblk}
\RequirePackage[colorlinks,citecolor=blue,urlcolor=blue]{hyperref}
\usepackage{graphicx}

\title{Modeling CD4$^+$ T cells dynamics in HIV-infected patients receiving repeated cycles of exogenous Interleukin 7}
\date{}

\author[1]{Ana Jarne}
\author[1]{Daniel Commenges}
\author[2]{M\'{e}lanie Prague}
\author[3,4]{Yves Levy}
\author[1,4]{Rodolphe Thi\'{e}baut}
\author[ ]{for inspire 2$\&$3 study group}
\affil[1]{INSERM-INRIA SISTM team}
\affil[2]{Harvard T.H. Chan School of Public Health}
\affil[3]{INSERM  U955}
\affil[4]{VRI Paris}

\begin{document}

\maketitle

\begin{abstract}
Combination Antiretroviral Therapy (cART) succeeds to control viral replication in most HIV infected patients. This is normally followed by a reconstitution of the CD4$^+$ T cells pool; however, this does not happen for a substantial proportion of patients. For these patients, an immunotherapy based on injections of Interleukin 7 (IL-7) has been recently proposed as a co-adjutant treatment in the hope of obtaining long-term reconstitution of the T cells pool. Several questions arise as to the long-term efficiency of this treatment and the best protocol to apply. Mathematical and statistical models can help answering these questions.

We develop a model based on a system of ordinary differential equations and a statistical model of variability and measurement. We can estimate key parameters of this model using the data from the main studies for this treatment, the INSPIRE, INSPIRE 2 $\&$ INSPIRE 3 trials. In all three studies, cycles of three injections have been administered; in the last two studies, for the first time, repeated cycles of exogenous IL-7 have been administered. Repeated measures of total CD4$^+$ T cells count in 128 patients as well as CD4$^+$Ki67$^+$ T cells count (the number of cells expressing the proliferation marker Ki67) in some of them were available. Our aim was to estimate the possible different effects of successive injections in a cycle, to estimate the effect of repeated cycles and to assess different protocols.

The use of dynamical models together with our complex statistical approach allow us to analyze major biological questions. We found a strong effect of IL-7 injections on the proliferation rate; however, the effect of the third injection of the cycle appears to be much weaker than the first ones. Also, despite a slightly weaker effect of repeated cycles with respect to the initial one, our simulations show the ability of this treatment of maintaining adequate CD4$^+$ T cells count for years. We were also able to compare different protocols, showing that cycles of two injections should be sufficient in most cases. 

\end{abstract}


\section{Introduction}

Infection by the Human Immunodeficiency Virus (HIV) leads to severe lymphopenia and general immune dysfunction. Combination Antiretroviral Therapy (cART) allows controlling viral load in most patients and often leads to an adequate immune restoration. However, not all patients get a satisfactory immune reconstitution despite undetectable viral load. Sereti et al [1] called these patients ``immunological non responders''; we prefer to call them ``immunological low responders'' because these patients can still experience an increase of CD4$^+$ T cells count under cART, albeit insufficient.

A treatment based on injections of exogenous Interleukin-7 (IL-7) has been recently proposed, and is for the moment the only promising approach in this context (Sereti et al [2], Levy et al [3,4]).  Endogenous IL-7 is a cytokine produced by non-marrow-derived stromal and epithelial cells, and since it was discovered in 1988 (Namen et al [5]), it has been found to play an important role in peripheral maintenance of T cells (Fry et al [6], Mackall et al [7]). In HIV-infected patients, a correlation between plasma levels of endogenous IL-7 and CD4$^+$ T cell counts has already been reported (Beq et al [8]), and different mechanisms of action of IL-7 regarding regulation of T lymphocytes number and behavior have been uncovered, as enhancing thymopoiesis (Mackall et al [7], Okamoto et al [9]), proliferation (Vieira et al [10], Sportes et al [11]) and survival (Seddon et al [12], Kondrack et al [13]) of CD4$^+$ T cells.

Mathematical representations of the behavior of the immune system in the context of HIV infection have been useful to describe and quantify biological processes that are not directly observed; the interaction between HIV virions and CD4$^+$ T cells was firstly modeled by Ho et al [14] and Perelson et al [15]. For modeling the effect of exogenous IL-7 administration, it is not useful to model virus concentration (because viral load is undetectable under cART), but it is necessary to distinguish between quiescent and proliferating cells. In this context, Thi\'{e}baut et al [16] have quantified the contribution of several biological mechanisms in CD4$^+$ T cells homeostasis. They have studied the effect of a single cycle of exogenous IL-7.
Here, we extend this approach with a modified statistical model for analyzing repeated cycles, based on data from 3 clinical studies, INSPIRE, INSPIRE 2 and INSPIRE 3.  We focus on several major clinical questions. What is the effect of the different injections in a cycle? What is the effect of repeated cycles? What is the long-term efficacy of this therapy in maintaining CD4$^+$ T cells count at a  satisfactory level (over 500 cells/$\mu$L)? What is the best protocol of injections?

This paper is divided into 7 sections. Section 2 gives an overview of the INSPIRE studies and the available data. Section 3 describes the main structure of the mathematical and statistical models. Section 4 presents and compares different statistical models: the ``basic model" studying the effect of exogenous IL-7 over a cycle as a whole, the ``3 $\beta$'s model" allowing the successive injections of a cycle to have different effects, and the ``cycle effect model" investigating the long term effect when administering repeated cycles. Section 5 compares results of four possible protocols (varying the number of injections of a cycle) and predicts their impact on the maintenance of CD4$^+$ T cells count $>$ 500 cells/$\mu$L for an average patient. Section 6 explores the possibility of optimizing the protocol by investigating in simulation the trajectories of CD4$^+$ T cells count in good and bad responders. Section 7 concludes.

\section{Data and materials}

\subsection{Data source and subjects}
The data have been compiled from three phase I/II multicenter studies: INSPIRE (Levy et al [4]), INSPIRE 2 and INSPIRE 3 (Thi\'{e}baut et al [17]). These studies investigated the effect of a \emph{purified glycosylated recombinant human Interleukin 7} (r-hIL-7) treatment on immune restoration in immunological low responder patients.
All participants were aged $\ge$ 18 years, were under stable cART for at least 2 years, presenting CD4$^+$ T cells count between 100-350 cells/$\mu$L (100-400 cells/$\mu$L for INSPIRE 2), and undetectable viral load for at least 6 months prior to screening.

In the first study, INSPIRE, 21 patients received three weekly injections (a ``complete cycle") of r-hIL-7 at different weight-dependent doses: 10, 20 and 30 $\mu$g/kg and the main objective was to evaluate the safety of this treatment.
INSPIRE 2 and INSPIRE 3 (with 23 and 84 treated patients, respectively) further studied the biological activity (as well as the safety) of repeated cycles of r-hIL-7 at 20 $\mu$g/kg. In this paper, data for all treated patients from the three studies (N=128) have been included from the time of the first injection.
Overall, 197 r-hIL-7 cycles were administered (41 of them were incomplete cycles consisting of 1 or 2 injections). More details are provided in a previous publication (Thi\'{e}baut et al [17]).

\subsection{Study design and observations}
Within the first INSPIRE study, all patients received complete cycles. They had clinic visits at weeks 1, 2 and 3 (at the moment of the injections), weeks 4, 5, 6, 9 and 12, and after, one visit every 3 months; see Levy et al [4] for more information. Among many measured biomarkers, our model uses total CD4$^+$ T cells count and the number of CD4$^+$ T cells expressing the Ki67 proliferation marker, hereafter called ``CD4 count" and ``Ki67 count", respectively. Measurements of CD4 counts were made at each visit, while Ki67 counts were only measured at weeks 1, 2, 3, 5 and 12.

For the first twelve patients of INSPIRE 2, clinic visits within the initial cycles were scheduled as for the INSPIRE study (for the rest of them, visits at week 9 were not performed). After, if CD4 counts were found to be below 550 cells/$\mu$L in one of the quarterly visits, a new r-hIL-7 cycle was administered (with the exception of the first 12 patients, who wait a year before receiving a new cycle). Within these repeated cycles, clinic visits were scheduled at weeks 1, 2 and 3 (at the moment of the injections), weeks 5 and 12, and once again quarterly visits are made to check the CD4 count. A maximum of 4 cycles within 21 months and a maximum of 3 cycles within 12 months were established, and all patients have been followed up at least 3 months after the last cycle. CD4 counts were measured at all visits for all patients, while Ki67 counts were measured only for the first cycles of the first 12 patients at weeks 1, 2, 3, 5 and 12.

For INSPIRE 3, patients were randomized into two arms: ``r-hIL-7 arm" and ``Control arm" with a ratio 3:1 (3 r-hIL-7 : 1 Control). Patients of the ``r-hIL-7 arm" received the same treatment scheme as patients from INSPIRE 2. Patients of the ``Control arm" were first followed up without receiving the r-hIL-7 for one year, and if CD4 count was still below 500 cells/$\mu$L, r-hIL-7 treatment was started as for the other group (Thi\'{e}baut et al [17]). CD4 counts were measured at all visits. No Ki67 counts measurements were available.

The total duration of the studies was 12, 24  and 21 months for INSPIRE, INSPIRE 2 and INSPIRE 3, respectively.

\section{Mathematical and statistical structure}

\subsection{Mathematical and statistical models}

Our theoretical framework to describe the dynamics of CD4 and Ki67 counts is based on the same system of ordinary differential equations (ODE) as proposed by Thi\'{e}baut et al [16]. For patient $i$ this model can be written as:
\begin{equation*}
\left\{
\begin{array}{lllllll}
\frac{dQ^i}{dt} & = & \lambda^i + 2 \rho^i P^i- \pi^i Q^i - \mu_Q^i Q^i\tabularnewline[0.2cm]
\frac{dP^i}{dt} & = & \pi^i Q^i - \rho^i P^i - \mu_P^i P^i
\end{array}
\right.
\end{equation*}

The initial condition is assumed to be the equilibrium point (specified by $\frac{dQ^i}{dt}(0) = 0 \; \text{,} \frac{dP^i}{dt}(0) = 0)$.

\begin{figure}[here]
 \centering\includegraphics[width=0.5\textwidth]{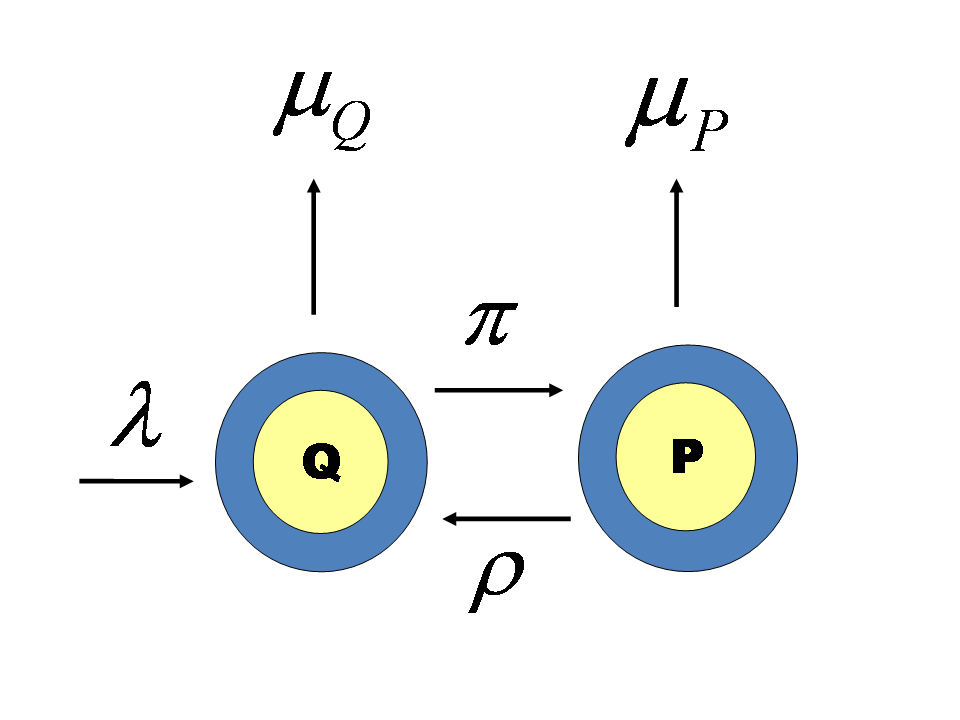}
 \vspace{-0.5cm}

 \caption{Graphical representation of the mathematical model}
 \label{fig:2comps}
 \end{figure}

A graphical representation of the system can be found in Figure \ref{fig:2comps}. This model includes two state variables: P, the concentration of proliferating cells expressing the Ki67 proliferation marker (CD4$^+$Ki67$^+$) and Q, the concentration of quiescent cells (CD4$^+$Ki67$^-$). We have also investigated a model with a feedback term, obtained by multiplying the basic proliferation rate by $\frac{1}{(P^i+Q^i)^{\nu}}$, where $\nu$ is a parameter to be estimated. We did not retain this feedback term because it did not lead to major improvement of the fit while requiring much more computation time (see Appendix A).

The vector of parameters of the ODE system is $\boldsymbol{\xi^i} = [\lambda^i,\rho^i,\pi^i,\mu_Q^i,\mu_P^i]$.
These parameters have a biological interpretation: $\lambda$ is the production rate, $\rho$ is the reversion rate, $\pi$ is the proliferation rate and $\mu_Q$ and $\mu_P$ are the mortality rates of Q and P cells, respectively. The logarithmic transformation ensures positivity of these biological parameters: $\boldsymbol{\tilde\xi}^i = \text{log} (\boldsymbol{\xi}^i)$.

Modeling the variability of the parameters is a crucial ingredient in our model because it allows to have a joint estimation of parameters across the population instead of fitting the model patient-by-patient. A mixed-effect model can be assumed for each transformed parameter $l$, $l$ = 1,...,$p$ (here $p$ = 5):
$$\tilde\xi^i_l (t) = \phi_l + \beta_l^\top z^i_l (t) + u_l^i$$
where $\phi_l$ is the intercept, $\beta_l$ is a vector of regression coefficients, $\boldsymbol{z^i_l}$ is a vector of explanatory variables, and $u_l^i$ are random effects assumed to be independently and identically normally distributed. Thus, the parameters can vary between subjects, but also with time through the time-dependent explanatory variables. In practice, for parsimony, random effects and explanatory variables are included for a subset of the parameters.

In this paper, we present and discuss several of these variability models. The random effects have been applied on $\lambda$ and $\rho$:  $u^i_{\lambda} \sim \mathcal{N} (0, \sigma_{\lambda}^2)$, $u^i_{\rho} \sim \mathcal{N} (0, \sigma_{\rho}^2)$ for all the models. The explanatory variables used are functions of the dose and of the timing of the r-hIL-7 injections and are used to model the proliferation rate ($\pi$) and the mortality rate ($\mu_Q$). These choices are based on many trials and on previous results of the literature (as in Thi\'{e}baut et al [16]).

We also need a model for the observations. The state variables $(P^i(t),Q^i(t))$ are not directly observable; we only have discrete-time observations of some functions of the components of this vector. Let Y$_{1j}^i$ and Y$_{2k}^i$ be the CD4 count and the Ki67 count for patient i at time $t_{ij}$ and $t_{ik}$, respectively. The following observation scheme is assumed:

\begin{equation*}
\left\{
\begin{array}{lllll}
(\text{Y}_{1j}^i)^{0.25} & = & ( P^i(t_{ij}) + Q^i (t_{ij}) )^{0.25} & + & \epsilon_{i1j}\\
(\text{Y}_{2k}^i)^{0.25} & = &  P^i(t_{ik})^{0.25} & + & \epsilon_{i2k}
\end{array}
\right.
\end{equation*}
with independently normally distributed measurement errors: $\epsilon_{i1j} \sim \mathcal{N} ( 0, \sigma_{CD4}^2)$, $\epsilon_{i2k} \sim \mathcal{N} ( 0, \sigma_{P}^2)$. Note that the times of observations may be different for the two observed components; indeed there were less observations of Ki67 counts than of CD4 counts.

\subsection{Inference}
The vector $\theta$ to be estimated includes the intercepts of the biological parameters $(\phi_{\lambda},\phi_{\rho},\phi_{\pi},\phi_{\mu_Q},\phi_{\mu_P})$, the regression coefficients $(\beta_{\pi},\beta_{\mu_Q})$, the variances of the random effects $(\sigma_{\lambda},\sigma_{\rho})$ and the variances of the measurement errors $(\sigma_{CD4},\sigma_{P})$. As in Guedj et al [18], first the individual likelihoods given the random effects can be computed; then, the individual likelihoods are computed by integrating over the random effects via the adaptive Gaussian quadrature (Genz et al [19], Pinheiro et al [20]); the global log-likelihood is the sum of the individual log-likelihoods. The parameters can then in principle be estimated by maximum likelihood. However, due to identifiability problems, it is useful to adopt an approximate Bayesian approach, as in Drylewicz et al [21]. The prior distribution $\pi(\theta)$ allows incorporating prior knowledge taken from the literature. In such very complex models MCMC algorithm generally fail, so we use an approximate Bayesian inference, simpler than the INLA approach of Rue et al [22] which is also difficult to apply here. Bayes theorem gives
$$\log [P(\theta \mid Y)]) = L(\theta) + \log [\pi(\theta)] + C,$$
where $P(\theta \mid Y)$ is the posterior distribution, $L(\theta)$ is the log-likelihood and C is the normalization constant. The Bernstein-Von Mises theorem (Van der Vaart [23]) justifies a normal approximation of the posterior (NAP). The NAP can be computed by maximizing the penalized log-likelihood $L^P(\theta) = L(\theta) + log[\pi(\theta)]$ and computing the inverse of the Hessian of $-L^P(\theta)$, $H_{L^P}^{-1}$. Thus, the NAP is $\mathcal{N} (\tilde\theta, H_{L^P}^{-1}(\tilde\theta))$.

This computation can be achieved with the NIMROD program (Prague et al [24]) which uses the so-called RVS algorithm (Commenges et al [25]); parallel computing is implemented to achieve acceptable computation times. Other approaches have been proposed for fitting ODE-based models: Ramsay et al [26] proposed a penalized likelihood approach for the trajectories of the state variables circumventing the need of solving the ODE system, but this approach has also numerical issues in presence of random effects; Kuhn et al [27] have proposed the stochastic approximation expectation maximisation (SAEM) algorithm which can also be used for maximising a log-likelihood or a penalized log-likelihood. One advantage of the RVS algorithm is the possibility of computing a stringent stopping criterion. See Appendix B for details.

\subsection{Comparison of different models}

Here, we present more than one possible statistical model to describe the effect of r-hIL-7 on biological parameters. To compare them, apart direct likelihood comparison and individual fits, we use an approximate cross-validation criterion, LCVa, proposed by Commenges et al [28]. LCVa is an extension of Akaike criterion (AIC), similar to the General Information Criterion (GIC) (Konishi et al [29]) that corrects not only for the number of parameters but also for the penalization; LCVa is normalized on the number of observations (see Commenges et al [30,31] for further developments). This criterion is:
$$\text{LCVa =} -\text{n}^{-1} [\text{ L}(\tilde\theta) - \text{Trace} (\text{H}^{-1}_{L^P}(\tilde\theta) \text{H}_L(\tilde\theta)) ] \text{,}$$
where H$_L$ is the Hessian of minus the log-likelihood. Since LCVa estimates a ``risk'' (cross-entropy or Kullback-Leibler risk equivalently), the smaller the better. Differences in criteria values between two models  can be considered as ``large" beyond 0.1 when the response is univariate. However, when the response is multivariate, the threshold for considering a difference as ``large'' should be higher, because LCVa, as defined here,  is normalized on the number of subjects and does not take into account the number of observations per subject. 

\section{Main results}

\subsection{Basic model: A cycle as a whole entity}
Firstly, we are interested in estimating the global effect of the first cycle of r-hIL-7. To begin with, only first received cycles for each patient have been considered. As in Thi\'{e}baut et al [16] the effect of r-hIL-7 is considered to be dose-dependent. In our case, we have chosen a to consider a power of the dose (as is common in pharmacology), that was fixed as 0.25 par profile likelihood (that is, the fourth root of the dose).

The effect on proliferation $\pi$ is taken into account during 7 days (this time was also fixed by profile likelihood) after each injection. Besides, the effect on the mortality rate $\mu_Q$ is considered to be constant from two days after the first injection during twelve months, followed by a linear decrease during another twelve months. As already mentioned, random effects are added on the production rate $\lambda$ and the reversion rate $\rho$.
Let d$_i$ the dose received for patient i, and let N$^i_t$ the number of injections that patient $i$ has received until time $t$. The statistical description for this first model is as follows:

\begin{equation*}
\left\{
\begin{array}{lllllllll}
\tilde\pi^i\text{(t)}&=&\tilde\pi_0& && + & \beta_{\pi} \; \; \; \text{d}_i^{0.25} & \mathds{1}_{\{N^i_t - N^i_{t-7} = 1\}}  \\
\tilde\lambda^i\text{(t)}&=&\tilde\lambda_0& + &  \text{u}_{\lambda}^i & & &    \\
\tilde\mu_{Q}^i\text{(t)}&=&\tilde\mu_{Q_0}& &&+& \beta_{\mu_Q} \;  \text{d}_i^{0.25} & f(t)  \\
\tilde\rho^i\text{(t)}&= &\tilde\rho_0& + & \text{u}_{\rho}^i &  &&\\
\tilde\mu_{P}^i\text{(t)}&=&\tilde\mu_{P_0}& & &&&
\end{array}
\right.
\label{eq:mixed}
\end{equation*}
where $\mathds{1}_{\{N^i_t - N^i_{t-7} = 1\}}$ is an indicator function taking value 1 if an injection has been administrated in the last 7 days, and

\begin{equation}
f(t) =
\left\{
\begin{array}{lll}
1 & if & 2 < t \leq 360 \\
1 - (t -  360)/360  & if & 360 < t \leq 720 \\
0 & if & 720 < t
\end{array}
\right.
\label{eq:equa}
\end{equation}

Taking the same priors as Thi\'{e}baut et al [16], we ran the analysis with the NIMROD program. The results are displayed in Table \ref{table:first}; r-hIL-7 injections increase the proliferation rate ($\pi$) from 0.041 per day at baseline to 0.135 per day during 7 days after each injection (for the dose equal to 20 $\mu$g/kg). Also the estimated mortality rate of Q cells decreases from 0.104 per day at baseline to 0.072 during the first year after the treatment.

\begin{table}[h]
\caption{\label{table:first} Priors and estimated mean and standard deviation (sd) of all parameters (in logarithmic and natural scales) for the ``basic" model when considering only the first cycle for all patients from INSPIRE 1, 2 $\&$ 3; Penalized (P) and Non Penalized (NP) likelihoods, and LCVa criteria}
 \begin{tabular}{rrrrrrrr}
\hline
 &  \multicolumn{ 2}{c}{PRIOR} & \multicolumn{ 2}{c}{POSTERIOR}  & \multicolumn{ 2}{c}{POSTERIOR} \\
 & \multicolumn{ 2}{c}{(log-scale)} & \multicolumn{ 2}{c}{(log-scale)} &  \multicolumn{ 2}{c}{(natural-scale)}  \\
 & mean &  sd  &  mean &     sd &   mean &   sd  \\
 \hline
$\lambda         $&        1.000&     1.000&     2.967&     0.062&    \textbf{19.440}&     \emph{1.196} \\
$\rho            $&       0.000&     0.250&     0.680&     0.095&     \textbf{1.973}&     \emph{0.187} \\
$\pi             $&      -4.000&     1.000&    -3.185&     0.115&     \textbf{0.041}&     \emph{0.005} \\
$\mu_Q           $&      -3.600&     0.500&    -2.264&     0.073&     \textbf{0.104}&     \emph{0.008} \\
$\mu_P           $&      -2.500&     0.500&    -1.550&     0.202&     \textbf{0.212}&     \emph{0.043} \\
\hline
$\beta_{\pi}     $&   &  &  &  &    \textbf{0.997}&     \emph{0.058} \\
$\beta_{\mu_Q}   $&  &  &  & &   \textbf{-0.305}&     \emph{0.020} \\
\hline
$\sigma_{\lambda}$ &  &  & & &     \textbf{0.254}&     \emph{0.025}\\
$\sigma_{\rho}   $ &  &  &  & &    \textbf{0.534}&     \emph{0.096}\\
\hline
$\sigma_{CD4}    $ &  &  &  & &    \textbf{0.254}&     \emph{0.003}\\
$\sigma_{P}      $ &  &  &   & &   \textbf{0.299}&     \emph{0.023}\\
\hline
P likelihood & \multicolumn{6}{c}{\textbf{-338.7}} \\
NP likelihood & \multicolumn{6}{c}{\textbf{-327.4}} \\
LCVa & \multicolumn{6}{c}{\textbf{2.558}} \\
\hline
 \end{tabular} \\
\end{table}

\subsection{3 $\beta$'s model: A cycle as three different injections}
Here we focus on a major question: Have all the three injections the same quantitative effect on proliferation of CD4$^+$ T cells? Or, more accurately, what is the role of every single injection in the whole effect of a cycle? For this model too, we only consider the first received cycle for each patient.The statistical model for $\pi$ was:

\begin{equation*}
\tilde\pi^i \text{(t)}= \tilde\pi_0 + \sum\limits_{k=1}^3  \mathds{1}_{\{N_t^i=k\}} \; \beta_{\pi_k} \; d_i^{0.25} \; \mathds{1}_{\{N^i_t - N^i_{t-7} = 1\}}
\label{eq:3betas}
\end{equation*}

The results are displayed in Table \ref{table:betas}. The quantitative effects of the successive injections are not equal. They are all significantly different from zero; the first and second one are similar but the effect of the third one is considerably weaker. With this model there is a noticeable improvement with respect to the previous one (LCVa is equal to 2.136 vs 2.558).

\begin{table}[h]
\caption{\label{table:betas} Priors and estimated mean and standard deviation (sd) of all parameters (in logarithmic and natural scales) for the ``3 $\beta$'s" model when considering only the first cycle for all patients from INSPIRE 1, 2 $\&$ 3; Penalized (P) and Non Penalized (NP) likelihoods, and LCVa criteria}
 \begin{tabular}{rrrrrrrr}
\hline
 &   \multicolumn{ 2}{c}{PRIOR} & \multicolumn{ 2}{c}{POSTERIOR}  & \multicolumn{ 2}{c}{POSTERIOR} \\
 &  \multicolumn{ 2}{c}{(log-scale)} & \multicolumn{ 2}{c}{(log-scale)} &  \multicolumn{ 2}{c}{(natural-scale)}  \\
 &  mean &  sd  &  mean &     sd &   mean &   sd  \\
 \hline
$\lambda         $&      1.000&     1.000&     2.355&     0.087&    \textbf{10.541}&     \emph{0.920} \\
$\rho            $&       0.000&     0.250&     0.635&     0.102&     \textbf{1.887}&     \emph{0.192} \\
$\pi             $&      -4.000&     1.000&    -3.306&     0.125&     \textbf{0.037}&     \emph{0.005} \\
$\mu_Q           $&       -3.600&     0.500&    -2.617&     0.080&     \textbf{0.073}&     \emph{0.006} \\
$\mu_P           $&     -2.500&     0.500&    -2.187&     0.258&     \textbf{0.112}&     \emph{0.029} \\
\hline
$\beta_{\pi1}    $&      &  & & &     \textbf{1.155}&     \emph{0.079} \\
$\beta_{\pi2}    $&     &  &  & &    \textbf{1.120}&     \emph{0.081} \\
$\beta_{\pi3}    $&     &  & & &     \textbf{0.622}&     \emph{0.073}\\
$\beta_{\mu_Q}   $&      &  &  & &   \textbf{-0.239}&     \emph{0.022}\\
\hline
$\sigma_{\lambda}$ &      &  &  & &    \textbf{0.267}&     \emph{0.025} \\
$\sigma_{\rho}   $ &      &  &   & &   \textbf{0.575}&     \emph{0.108} \\
\hline
$\sigma_{CD4}    $ &  &  &  & &    \textbf{0.241}&     \emph{0.003} \\
$\sigma_{P}      $ &  &  &  & &    \textbf{0.305}&     \emph{0.025}\\
\hline
P likelihood & \multicolumn{6}{c}{\textbf{-279.8}} \\
NP likelihood & \multicolumn{6}{c}{\textbf{-273.3}} \\
LCVa & \multicolumn{6}{c}{\textbf{2.136}} \\
\hline
\end{tabular} \\
\end{table}

\subsection{Cycle effect model: Effect of successive cycles}
Among the 128 treated patients from all the three studies, 74 have received more than one cycle. A key question is: Have these repeated cycles the same quantitative effect with respect to initial ones? CD4 counts are higher before starting repeated cycles. Also, antibodies anti-r-hIL-7 could appear after an initial cycle, modifying the effect of r-hIL-7 when cycles are repeated. The second goal of this paper is to estimate possible quantitative differences in repeated versus initial cycles. To make this possible, we included data from all received cycles and we estimated a new fixed effect: the ``cycle effect" $\beta_C$. We keep the notation t$_{i1}$ for the time when patient $i$ receives the first injection of a cycle.
If C(t) counts the number of cycles received at time $t$, let $\mathds{1}_{C(t) > 1}$ be 1 if a cycle has been received before time $t$, 0 otherwise. The cycle effect is incorporated into the statistical model of proliferation rate as follows:

\begin{equation*}
\tilde\pi^i \text{(t)}= \tilde\pi_0 + \Big[ \beta_C \mathds{1}_{\{C(t) > 1\}} + \sum\limits_{k=1}^3  \mathds{1}_{\{N_t^i=k\}} \; \beta_{\pi_k} \; d_i^{0.25}\Big] \; \mathds{1}_{\{{N^i_t - N^i_{t-7}} = 1\}}
\label{eq:betacycle}
\end{equation*}

The results are displayed in Table \ref{table:cycle}. The posterior distribution of the cycle effect $\beta_C$ has mean equal to -0.163 and standard deviation equal to 0.015. In other words, the cycle effect is found to be significantly negative. In natural scale, the effect on proliferation rate for successive cycles is found to be e$^{-0.163}$ = 0.85 times the effect of the first cycle. The biological interpretation of the cycle effect is not yet clearly explained. One explanation may be that the first cycle has modified  the reaction of the immune system to further injections; one possibility is that antibodies against IL-7 decrease the efficient concentration of IL-7 obtained at the target. However, we must take into consideration differences in mean CD4 count before the initial and repeated cycles. The mean CD4 count at baseline was 266 cells/$\mu$L whereas it was 456 cells/$\mu$L before repeated cycles. Considering the homeostatic regulation of the population of CD4$^+$ cells, that prevents CD4 counts from exceeding 1200-1300 cells/$\mu$L, a feedback mechanism may explain an apparent cycle effect. With the aim to deeper study this phenomenon, we have incorporated a feedback term (see Appendix A). We found that a feedback effect could indeed be detected, but this had no major influence on the estimate of the cycle effect. 

\begin{table}[h]
\caption{\label{table:cycle} Priors and estimated mean and standard deviation (sd) of all parameters (in logarithmic and natural scales) for the ``cycle effect" model when considering all cycles for all patients from INSPIRE 1, 2 $\&$ 3; Penalized (P) and Non Penalized (NP) likelihoods, and LCVa criteria}
 \begin{tabular}{rrrrrrr}
\hline
 &  \multicolumn{ 2}{c}{PRIOR} & \multicolumn{ 2}{c}{POSTERIOR}  & \multicolumn{ 2}{c}{POSTERIOR} \\
 &  \multicolumn{ 2}{c}{(log-scale)} & \multicolumn{ 2}{c}{(log-scale)} &  \multicolumn{ 2}{c}{(natural-scale)}  \\
 &  mean &  sd  &  mean &     sd &   mean &   sd  \\
 \hline
$\lambda         $&   1.000&     1.000&     1.672&     0.061&     \textbf{5.323}&     \emph{0.326} \\
$\rho            $&    0.000&     0.250&     0.892&     0.093&     \textbf{2.440}&     \emph{0.226} \\
$\pi             $&   -4.000&     1.000&    -2.853&     0.074&     \textbf{0.058}&     \emph{0.004} \\
$\mu_Q           $&  -3.600&     0.500&    -2.610&     0.068&     \textbf{0.074}&     \emph{0.005} \\
$\mu_P           $&   -2.500&     0.500&    -2.567&     0.200&     \textbf{0.077}&     \emph{0.015} \\
\hline
$\beta_{\pi1}    $&  &  &    &  &   \textbf{0.931} &     \emph{0.042}\\
$\beta_{\pi2}    $&  &  &    &  &   \textbf{0.707} &     \emph{0.043}\\
$\beta_{\pi3}    $&  &  &   &  &    \textbf{0.229} &     \emph{0.042}\\
$\beta_{\mu_Q}   $&   &  &   &  &  \textbf{-0.082} &     \emph{0.006}\\
$\beta_{C}   $&  &  &   &  &  \textbf{-0.163} &     \emph{0.015}\\
\hline
$\sigma_{\lambda}$ &  &  &  &  &  \textbf{0.243} &     \emph{0.026}\\
$\sigma_{\rho}   $ &  &  &  &  &  \textbf{0.515} &     \emph{0.084}\\
\hline
$\sigma_{CD4}    $ &  &  &  &  &  \textbf{0.289} &     \emph{0.003}\\
$\sigma_{P}      $ &  &  &  &  & \textbf{0.281} &     \emph{0.019}\\
\hline
P likelihood & \multicolumn{6}{c}{\textbf{-618.6}} \\
NP likelihood & \multicolumn{6}{c}{\textbf{-609.4}} \\
LCVa & \multicolumn{6}{c}{\textbf{4.762}} \\
\hline
\end{tabular} \\
\end{table}

Appendix C and D show some fits of real data from INSPIRE 2 and 3 obtained with this model. Individual predicted trajectories were computed using the Parametric Empirical Bayes (PEB) for the parameters having a random effect ($\lambda$ and $\rho$). Several protocols have been compared in the next Section by means of this model.

\section{Comparing different protocols for an average patient}

We have used the ``cycle effect model" to compare different administration protocols of r-hIL-7 containing complete and incomplete cycles. We always assumed that CD4 counts are measured every three months, and a new cycle is administered when CD4 count $<$ 550 cells/$\mu$L for 4 years. We examined four possibilities: in protocol A, the patient always receives complete cycles; in protocol B, the patient receives a first complete cycle followed by repeated cycles composed of two injections; in protocol C the patient receives a first complete cycle followed by repeated cycles of one single injection; in protocol D the patient always receives 2-injection cycles (including the initial one).

The protocols were compared according to three criteria computed over a four-year period: number of injections and cycles received, median CD4 count over the follow-up and time spent below 500 cells/$\mu$L. The criteria were computed for an average patient having both random effects equal to zero (equilibrium values of 272 and 6.3 for CD4 and Ki67 counts, respectively).

\begin{figure}[!h]
 \includegraphics[width=1\textwidth]{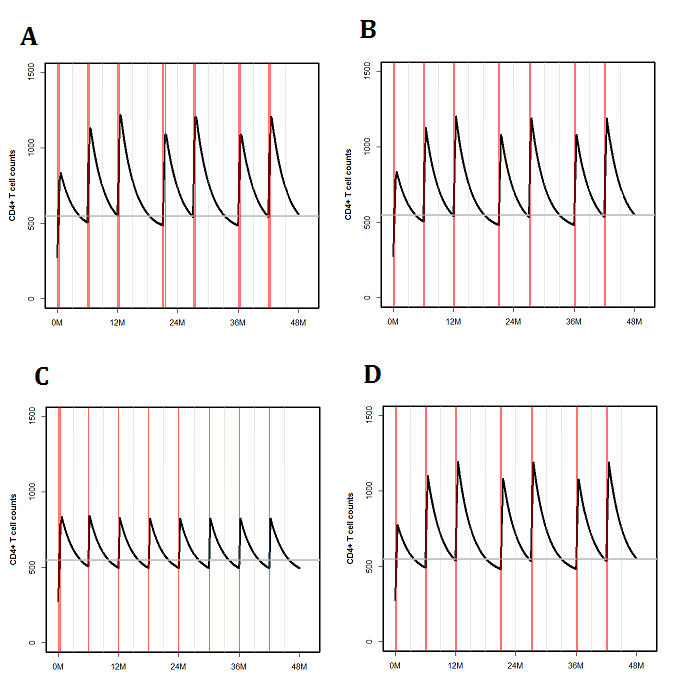}
 \caption{CD4 count (cells/$\mu$L) predictions for 4 years for a patient having b$_{\rho}^i$ = b$_{\lambda}^i$ = 0. Protocols A, B and C include a first complete cycle followed by: complete cycles (A), two-injection cycles (B) and one-injection cycles (C). Protocol D includes only 2-injection cycles. Vertical dotted lines are CD4 count assessments (every three months) and vertical solid lines are injections. Horizontal line marks the CD4 threshold of 550 cells/$\mu$L.}
 \label{pred}
 \end{figure}

The results are displayed in Table \ref{normal}, and we can observe the expected trajectories in Figure \ref{pred}.

A complete cycle followed by 2-injection cycles (Protocol B) could lead to similar results than Protocol A in terms of median CD4 count, with the non negligible advantage that Protocol B requires 15 injections instead of 21. Protocol C ensures and identical time spent under 500 cells/$\mu$L with only 10 injections, but achieves a median CD4 count lower than Protocol A. Protocol D is also slightly worse than Protocol A in terms of time below 500 cells/$\mu$L.

\begin{table}[!h]
\centering
\caption{\label{normal} Comparison of the number of injections and cycles received, time under 500 CD4 count and median CD4 count for a patient with RE equal to zero for the four protocols through four years. In protocol A, the patient always receives complete cycles; in protocol B, the patient receives a first complete cycle followed by repeated cycles composed of two injections; in protocol C the patient receives a first complete cycle followed by repeated cycles of one single injection; in protocol D the patient always receives 2-injection cycles (including the initial one)}
\vspace{0.5cm}

\begin{tabular}{ccccc}
     &  \large{\textbf{A}} &  \large{\textbf{B}} &  \large{\textbf{C}} &  \large{\textbf{D}} \\
     &  &  &  &  \\
\hline
     &  &  &  &  \\
Number of injections received &  21 & 15 & 10 & 14 \\
     &  &  &  &  \\
Number of cycles received & 7 & 7 & 8 & 7 \\
     &  &  &  &  \\
Time under 500 CD4/$\mu$L (days) & 60 & 73 & 60 & 87 \\
     &  &  &  &  \\
Median CD4 count & 678 & 663 & 588 & 654 \\
     &  &  &  &  \\
\hline
\end{tabular}
\end{table}

\section{Adaptive protocols: towards a personalized medicine}

Let a patient beginning the r-hIL-7 treatment with a first cycle during which we collect several CD4 count measurements in order to know the value of his random effects. This information can be used to calculate the expected trajectories of this patient when applying the 4 previous protocols. We have taken individual information for two real patients from INSPIRE 2 and INSPIRE 3 studies and we want to compare what would be the best protocol for them.

Firstly, we have chosen a patient having a very good response in terms of CD4 count. For this patient, the value of the parameters including random effects are equal to $\lambda$ = 6.586 and $\rho$ = 4.797 (all the other parameters are the population parameters obtained in the ``cycle effect" model).

\begin{figure}[!h]
 \includegraphics[width=1\textwidth]{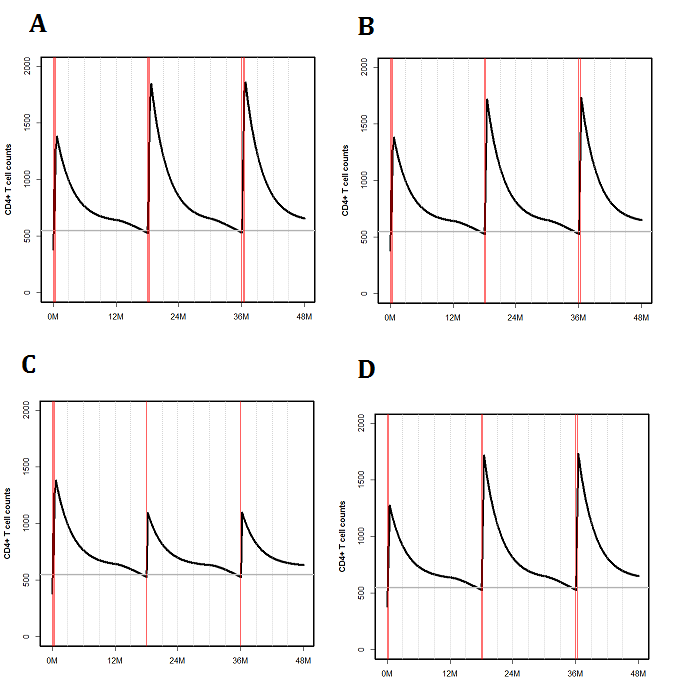}
 \caption{CD4 count (cells/$\mu$L) predictions for 4 years for a particularly good responder patient. Protocols A, B and C include a first complete cycle followed by: complete cycles (A), two-injection cycles (B) and one-injection cycles (C). Protocol D includes only 2-injection cycles. Vertical dotted lines are CD4 count assessments (every three months) and vertical solid lines are injections. Horizontal line marks the CD4 threshold of 550 cells/$\mu$L.}
 \label{bon}
 \end{figure}

In Figure \ref{bon} we predict the expected trajectories and Table \ref{good} displays the four criteria for this patient. According to our model, there are only minor differences between the four protocols for this patient for the four criteria. Protocol B would spare 2 injections with little impact on the CD4 count and even Protocol C would be admissible.

\begin{table}[!h]
\centering
\caption{\label{good} Comparison of the number of injections and cycles received, time under 500 CD4 count and median CD4 count for a "good responder" patient for the four protocols through four years. In protocol A, the patient always receives complete cycles; in protocol B, the patient receives a first complete cycle followed by repeated cycles composed of two injections; in protocol C the patient receives a first complete cycle followed by repeated cycles of one single injection; in protocol D the patient always receives 2-injection cycles (including the initial one)}
\vspace{0.5cm}

\begin{tabular}{ccccc}
     &  \large{\textbf{A}} &  \large{\textbf{B}} &  \large{\textbf{C}} &  \large{\textbf{D}} \\
     &  &  &  &  \\
\hline
     &  &  &  &  \\
Number of injections received &   9 &  7 &  5 &  6 \\
     &  &  &  &  \\
Number of cycles received &   3 &  3 &  3 & 3 \\
     &  &  &  &  \\	
Time under 500 CD4/$\mu$L (days) &  3 & \centering 3 &  3 & g 3 \\
     &  &  &  &  \\
Median CD4 count &  721 &  709 & 669 &  703 \\
     &  &  &  &  \\
\hline
\end{tabular}
\end{table}

We have chosen another patient having a particularly poor response to the r-hIL-7 treatment. In this case, the value of the parameters including random effects is equal to $\lambda$ = 3.284 and $\rho$ = 1.956.

\begin{figure}[!h]
 \includegraphics[width=1\textwidth]{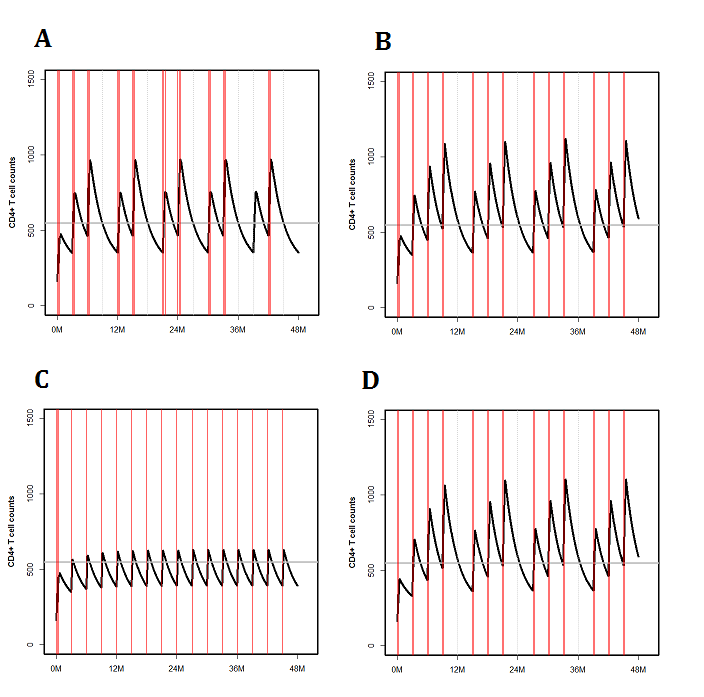}
 \caption{CD4 count (cells/$\mu$L) predictions for 4 years for a patient with a particularly poor response. Protocols A, B and C include a first complete cycle followed by: complete cycles (A), two-injection cycles (B) and one-injection cycles (C). Protocol D includes only 2-injection cycles. Vertical dotted lines are CD4 count assessments (every three months) and vertical solid lines are injections. Horizontal line marks the CD4 threshold of 550 cells/$\mu$L.}
 \label{bad}
 \end{figure}

Figure \ref{bad} displays the expected trajectories for the different protocols and Table \ref{mauvais} gives the four criteria for this patient. Our model predicts that this patient could benefit from 2-injection cycles (protocol B) without loss of efficiency in terms of CD4 count or time over 500 cells/$\mu$L. However, 1-injection cycles (Protocol C) would not be enough.

\begin{table}[!h]
\centering
\caption{\label{mauvais} Comparison of the number of injections and cycles received, time under 500 CD4 count and median CD4 count for a "bad responder" patient for the four protocols through four years. In protocol A, the patient always receives complete cycles; in protocol B, the patient receives a first complete cycle followed by repeated cycles composed of two injections; in protocol C the patient receives a first complete cycle followed by repeated cycles of one single injection; in protocol D the patient always receives 2-injection cycles (including the initial one)}
\vspace{0.5cm}

\begin{tabular}{ccccc}
     &  \large{\textbf{A}} &  \large{\textbf{B}} &  \large{\textbf{C}} &  \large{\textbf{D}} \\
     &  &  &  &  \\
\hline
     &  &  &  &  \\
Number of injections received &  30 & 27 & 18 & 26 \\
     &  &  &  &  \\
Number of cycles received & 10 & 13 & 16 & 13 \\		
     &  &  &  &  \\		
Time under 500 CD4/$\mu$L (days) & 561 & 366 & 893 & 381 \\
     &  &  &  &  \\
Median CD4 count & 549 & 617 & 470 & 611 \\
     &  &  &  &  \\
\hline
\end{tabular}
\end{table}

\section{Discussion}
INSPIRE 2 and INSPIRE 3 are the first studies where repeated cycles of r-hIL-7 were administrated to test the long-term restoration of the immune system in low immunological responders. Here we have used a simple mathematical model with complex statistical approaches to model the effect of these repeated cycles on CD4$^+$ T cells concentration. We worked with two CD4$^+$ T cells populations: quiescent and proliferating (presenting the Ki67$^+$ marker).

When considering every injection separately, the first important result of this paper is that our model predicts a decreasing effect of successive injections on proliferation rate; the third injection seems to have a weaker effect. We also found that the effect of repeated cycles on proliferation rate was slightly weaker than the effect of the initial one; the order of magnitude, however, is the same. This can be due to the natural homeostatic regulation of CD4$^+$ T cells, since repeated cycles start at a higher CD4 count. In order to investigate this question, we have introduced a feedback term; in this case the feedback term slightly improved the fit but the estimate of the ``cycle effect" did not change much. Thus although a feedback mechanism is plausible, there may be other reason, such as the presence of antibodies,  for a slightly weaker effect of repeated cycles.

Simulations show how these repeated cycles are able to maintain adequate CD4 counts for a long time. We have compared four protocols and shown that cycles of two injections should be sufficient, sparing a certain number of injections without detrimental effect on CD4 count. Our results agree with a survival analysis presented in Thi\'{e}baut et al [17] who compared the time spent over 500 cells/$\mu$L after a 3-injection cycle and a 2-injection cycle.

Also, the inclusion of random effects is a key ingredient when considering dynamic models as assistance for treatment personalized decisions. Inter-individual differences in parameters imply inter-individual differences in expected trajectories that can be used for devising adaptive treatment strategies (Prague et al [32]). We could use this mechanistic model for guiding the treatment, with the aim of minimizing the number of administered injections within repeated cycles ensuring the expected response. Predictions could also easily be made for different time lapses between cycles or thresholds for receiving a new cycle.

Some other questions regarding the interaction between the r-hIL-7 and the immune system could be modeled with additional data. For instance, preferential effects on specific T cell subsets as recent thymic emigrants (RTEs) and naive non-RTE T cell populations (Mackall et al [33]) could be analyzed.

\clearpage

\section*{APPENDIX A: Model with a feedback term}

Trajectories satisfying an ODE system have an intrinsic tendency to return to the equilibrium point, when it exists, which is the case for the systems proposed in this paper. In this sense, a feedback term is not necessary to ensure homeostasis, a key concept in physiology. We have, however, considered adding a feedback term in the mathematical model in order to examine the cycle effect $\beta_C$ in depth. This term will explicitly avoid CD4$^+$ T cells to proliferate without control and possibly ensure a faster return to an equilibrium point. The simplest feedback term is $[\frac{1}{P+Q}]^{\nu}$, and can be added in both equations to the proliferation term. The system with feedback is as follows:

$$
\left\{
\begin{array}{l}
 \frac{dQ^i}{dt} = \lambda^i + 2 \rho^i P^i - \mu_Q^i Q^i - \pi^i Q^i \boldsymbol{\frac{1}{(P^i+Q^i)^{\nu}}}
\\
\\
 \frac{dP^i}{dt} = \pi^i Q^i \boldsymbol{\frac{1}{(P^i+Q^i)^{\nu}}} - \rho^i P^i - \mu_P^i P^i
\end{array}
\right.
$$

Models with feedback were fitted using the 39 patients of INSPIRE who had Ki67 count measurements. The feedback coefficient was estimated at $\nu =  0.119$.
In Table \ref{FB} we compare some models with and without feedback term.

\begin{table}[h]
\caption{\label{FB} Comparison of loglikelihoods and LCVa criteria of models with and without feedback for all INSPIRE patients with CD4 and Ki67 count measurements (N=39)}
\begin{tabular}{c|cc|cc|cc}
\hline
 & \multicolumn{2}{c}{WITHOUT FEEDBACK}  & \multicolumn{2}{c}{WITH FEEDBACK} \\
 & Basic model & 3 $\beta$'s model & Basic model & 3 $\beta$'s model \\
\hline
NP loglike & -44.643 & -36.549 & -41.735 & -36.419\\
P loglike & -49.393 & -41.306 & -46.965 & -41.015\\
LCVa & 1.146 & 0.940 & 1.073 & 0.963\\
\hline
\end{tabular}
\end{table}

The feedback term does not lead to a great improvement of the LCVa criterion, especially for the 3 $\beta$'s model. 
 
The detection of a cycle effect raises anew the issue of a possible feedback. It may be that the feedback could not be detected when starting with very low CD4 count, but could be more visible when starting at higher CD4 count; this feedback might explain the apparent cycle effect. To answer this question we ran the model for repeated cycles with feedback. With this more complicated model and larger data set, we could not directly estimate the parameter $\nu$, so we resort to profile likelihood. Computing the likelihood for $\nu=0.05, 0.1, 0.15,0.20,0.25,0.30$ we found that the best likelihood was obtained for $\nu=0.1$, a value close to what was estimated in the small data set ($\nu =  0.119$). 
 The results are shown in Table \ref{table:FB}.

\begin{table}[h]
\caption{\label{table:FB} Priors and estimated mean and standard deviation (sd) of all parameters (in logarithmic and natural scales) for the ``cycle effect" model when considering all cycles for each patient including a feedback term with $\nu$ = 0.1; Penalized (P) and Non Penalized (NP) likelihood and LCVa criteria}
 \begin{tabular}{rrrrrrrr}
\hline
 &  \multicolumn{ 2}{c}{PRIOR} & \multicolumn{ 2}{c}{POSTERIOR}  & \multicolumn{ 2}{c}{POSTERIOR} \\
 & \multicolumn{ 2}{c}{(log-scale)} & \multicolumn{ 2}{c}{(log-scale)} &  \multicolumn{ 2}{c}{(natural-scale)}  \\
 & mean &  sd  &  mean &     sd &   mean &   sd  \\
 \hline
$\lambda         $& 1.000&     1.000&     0.275&     0.157&     1.316&     0.207 \\
$\rho            $&  0.000&     0.250&     1.052&     0.083&     2.863&     0.238 \\
$\pi             $&  -4.000&     1.000&    -1.975&     0.068&     0.139&     0.009 \\
$\mu_Q           $& -3.600&     0.500&    -2.538&     0.067&     0.079&     0.005 \\
$\mu_P           $&  -2.500&     0.500&    -2.212&     0.138&     0.109&     0.015 \\
\hline
$\beta_{\pi1}    $&  &  & & &    0.806&     0.038\\
$\beta_{\pi2}    $&  &  & & &    0.626&     0.037\\
$\beta_{\pi3}    $&  &  &  & &   0.212&     0.035\\
$\beta_{\mu_Q}   $&  &  &  & &  -0.063&     0.005\\
$\beta_{C}   $&  &  &  & &  -0.153&     0.015\\
\hline
$\sigma_{\lambda}$ &  &  & & &   -0.608&     0.097 \\
$\sigma_{\rho}   $ &  &  &  & &  -0.440&     0.071\\
\hline
$\sigma_{CD4}    $ &  &  &  & &     0.286&     0.004 \\
$\sigma_{P}      $ &  &  & & &    0.301&     0.021\\
\hline
P likelihood & \multicolumn{6}{c}{\textbf{-598.0}} \\
NP likelihood & \multicolumn{6}{c}{\textbf{-584.5}} \\
LCVa & \multicolumn{6}{c}{\textbf{4.567}} \\
\hline
\end{tabular}
\end{table}

For the repeated cycles data set, the feedback term leads to an improvement of the LCVa criterion. This may reflect a biological feedback mechanism. However, this does not modify the cycle effect $\beta_C$. 

\clearpage

\section*{APPENDIX B: Identifiability and convergence}\label{Identifiability}

As can be easily verified, both models with and without the feedback term, present no problems regarding the ``theoretical" identifiability (that depends on the model structure) but even so, they could present ``practical" identifiability problems as explained in Guedj et al [18]. In fact, practical identifiability problems are a mix of statistical and numerical problems which are difficult to disentangle; with scarce information, the variances of the estimators are large, but it comes also with a flat shape of the log-likelihood, making it difficult to maximize. The difficulty is enhanced by the fact that there are several layers of numerical computation needed to compute the likelihood, leading to an accumulation of numerical errors.

A crucial point in an iterative algorithm is the stopping criteria. Besides the displacement in the parameter space and the variation of the likelihood function, another convergence criterion proposed by Commenges et al [25] has been implemented in NIMROD. It is the Relative Distance to Maximum (RDM) defined as
$$\text{RDM}(\theta^{(k)}) = \frac{U^P(\theta^{(k)})^T G^{-1}(\theta^{(k)}) U^P(\theta^{(k)})}{p}$$
where $U^P(\cdot)$ is the penalized score and $G(\cdot)$ is an approximation of the Hessian of minus the penalized likelihood. This criterion can be interpreted as the ratio of the numerical error over the statistical error, and is asymptotically invariant near the maximum to any one-to-one transformation of the parameters. Prague et al [24] propose 0.1 as a good default value.

\clearpage

\section*{ APPENDIX C: Some fits of total CD4$^+$ T cell counts}\label{fitsp+q}

\begin{figure}[h!]
\includegraphics[width=12cm]{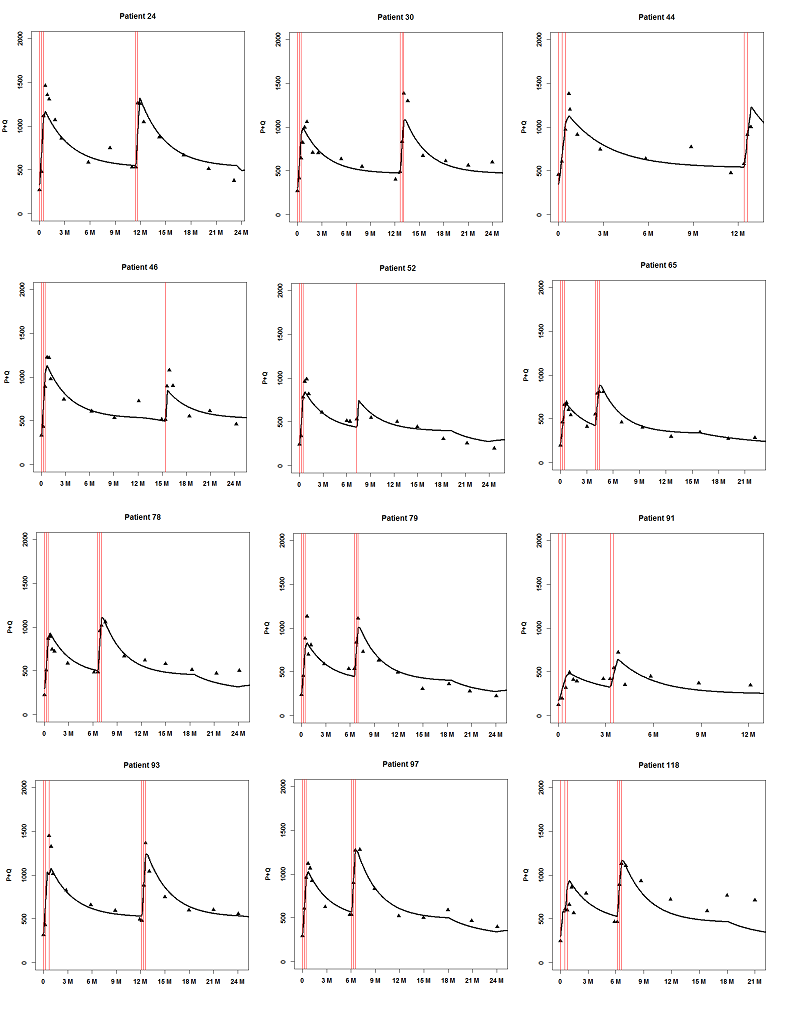}
\caption{ \emph{Cycle effect model}: Fits of total CD4 count for 12 patients from INSPIRE 2 and 3 chosen randomly among those who received more than a cycle.}

\end{figure}

\clearpage

\section*{APPENDIX D: Some fits of CD4$^+$Ki67$^+$ T cells}\label{fitsp}

\begin{figure}[h!]
\includegraphics[width=12cm]{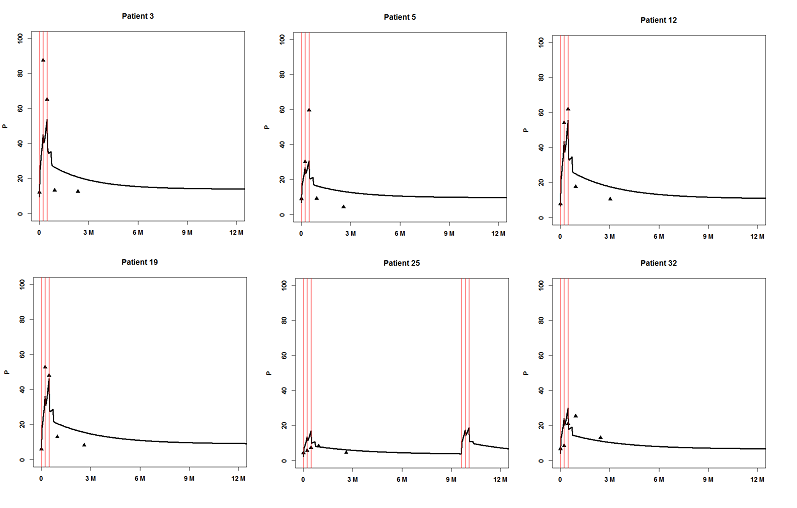}
\caption{\emph{Cycle effect model}: Fits of Ki67 count for 6 patients from INSPIRE and INSPIRE 2 chosen randomly among those who had measurements for this biomarker (only during the first cycle).}
\end{figure}

\clearpage

\section*{Acknowledgements}

We would like to thank the main investigators and supervisors of INSPIRE 2 and 3 studies: Jean-Pierre Routy, Irini Sereti, Margaret Fischl, Prudence Ive, Roberto F. Speck, Gianpiero D'Offizi, Salvatore Casari, Sharne Foulkes, Ven Natarajan, Guiseppe Tambussi, Michael M. Lederman, Th\'{e}rese Croughs and Jean-François Delfraissy. We have benefited greatly from the computing facilities MCIA (M\'{e}socentre de Calcul Intensif Aquitain) of the Universit\'{e} de Bordeaux and of the Universit\'{e} de Pau et des Pays de l'Adour for parallel computing. Also, grateful acknowledgement is made to the Vaccine Research Institute for grant support. 

\vspace{1cm}

\section*{References}
\vspace{0.2cm}

\hspace{0.5cm} [1] Sereti, I., Estes, J. D., Thompson, W. L., Morcock, D. R., Fischl, M. A., Croughs, T., Beq, S., de Micheaux, S. L., Yao, M. D., Ober, A. et al. (2014).
Decreases in colonic and systemic inflammation in chronic HIV infection after IL-7 administration. \emph{PLoS Pathog} \textbf{10} e1003890.
\vspace{0.2cm}

[2] Sereti, I., Dunham, R. M., Spritzler, J., Aga, E., Proschan, M. A., Medvik, K., Battaglia, C. A., Landay, A. L., Pahwa, S., Fischl, M. A. et al. (2009). 
IL-7 administration drives T cell cycle entry and expansion in HIV-1 infection. \emph{Blood} \textbf{113} 6304-6314.
\vspace{0.2cm}

[3] Levy, Y., Lacabaratz, C., Weiss, L., Viard, J.P., Goujard, C., Leli\`{e}vre, J.D., Bou\'{e}, F., Molina, J.M., Rouzioux, C., Avettand-F\'{e}no\^{e}l, V. et al. (2009).
Enhanced T cell recovery in HIV-1-infected adults through IL-7 treatment. \emph{The Journal of clinical investigation} \textbf{119} 997.
\vspace{0.2cm}

[4] Levy, Y., Sereti, I., Tambussi, G., Routy, J., Lelievre, J., Delfraissy, J., Molina, J., Fischl, M., Goujard, C., Rodriguez, B. et al. (2012). 
Effects of recombinant human interleukin 7 on T-cell recovery and thymic output in HIV-infected patients receiving antiretroviral therapy: 
results of a phase I/IIa randomized, placebo controlled, multicenter study. \emph{Clinical infectious diseases} \textbf{55} 291-300.
\vspace{0.2cm}

[5] Namen, A., Schmierer, A., March, C., Overell, R., Park, L., Urdal, D. and Mochizuki, D. (1988). 
B cell precursor growth-promoting activity. Purication and characterization of a growth factor active on lymphocyte precursors.
\emph{The Journal of experimental medicine} \textbf{167} 988-1002.
\vspace{0.2cm}

[6] Fry, T. J. and Mackall, C. L. (2002). Interleukin-7: from bench to clinic. \emph{Blood} \textbf{99} 3892-3904.
\vspace{0.2cm}

[7] Mackall, C.L., Fry, T.J., Bare, C., Morgan, P., Galbraith, A. and Gress, R.E. (2001). IL-7 increases both thymic-dependent and thymic-independent T-cell 
regeneration after bone marrow transplantation. \emph{Blood} \textbf{97} 1491-1497.
\vspace{0.2cm}

[8] Beq, S., Rannou, M.T., Fontanet, A., Delfraissy, J.F., Th\`{e}ze, J. and Colle, J.H. (2004). 
HIV infection: pre-highly active antiretroviral therapy IL-7 plasma levels correlate with long-term CD4 cell count increase after treatment. \emph{Aids} \textbf{18} 563-565.
\vspace{0.2cm}

[9] Okamoto, Y., Douek, D.C., McFarland, R.D. and Koup, R.A. (2002). Effects of exogenous interleukin-7 on human thymus function. \emph{Blood} \textbf{99} 2851-2858.
\vspace{0.2cm}

[10] Vieira, M., Soares, D., Borthwick, N.J., Maini, M.K., Janossy, G., Salmon, M. and Akbar, A.N. (1998). 
IL-7-dependent extrathymic expansion of CD45RA+ T cells enables preservation of a naive repertoire. \emph{The Journal of Immunology} \textbf{161} 5909-5917.
\vspace{0.2cm}

[11] Sport\`{e}s, C., Hakim, F.T., Memon, S.A., Zhang, H., Chua, K.S., Brown, M.R., Fleisher, T.A., Krumlauf, M.C., Babb, R.R., Chow, C.K. et al. (2008). 
Administration of rhIL-7 in humans increases in vivo TCR repertoire diversity by preferential expansion of naive T cell subsets. \emph{The Journal of experimental medicine} \textbf{205} 1701-1714.
\vspace{0.2cm}

[12] Seddon, B., Tomlinson, P. and Zamoyska, R. (2003). Interleukin 7 and T cell receptor signals regulate homeostasis of CD4 memory cells. \emph{Nature immunology} \textbf{4} 680-686.
\vspace{0.2cm}

[13] Kondrack, R.M., Harbertson, J., Tan, J.T., McBreen, M.E., Surh, C.D. and Bradley, L.M. (2003). 
Interleukin 7 regulates the survival and generation of memory CD4 cells. \emph{The Journal of experimental medicine} \textbf{198} 1797-1806.
\vspace{0.2cm}

[14] Ho, D.D., Neumann, A.U., Perelson, A.S., Chen, W., Leonard, J.M., Markowitz, M. et al. (1995). 
Rapid turnover of plasma virions and CD4 lymphocytes in HIV-1 infection. \emph{Nature} \textbf{373} 123-126.
\vspace{0.2cm}

[15] Perelson, A.S., Neumann, A.U., Markowitz, M., Leonard, J.M. and Ho, D.D. (1996). 
HIV-1 dynamics in vivo: virion clearance rate, infected cell life-span, and viral generation time. \emph{Science} \textbf{271} 1582-1586.
\vspace{0.2cm}

[16] Thi\'{e}baut, R., Drylewicz, J., Prague, M., Lacabaratz, C., Beq, S., Jarne, A., Croughs, T., Sekaly, R.P., Lederman, M.M., Sereti, I. et al. (2014). 
Quantifying and Predicting the Effect of Exogenous Interleukin-7 on CD4+ T Cells in HIV-1 Infection. \emph{PLoS computational biology} 10 e1003630.
\vspace{0.2cm}

[17] Thi\'{e}baut, R., Jarne, A., Routy, J.P., Sereti, I., Fischl, M., Ive, P., Speck, R.F., D\'Offizi, G., Casari, S., Commenges, D., Foulkes, S., Natarajan, V., Croughs, T., Delfraissy, J.F.,  Tambussi, G., Levy, Y., Lederman, M.M. (2016). 
Repeated Cycles of Recombinant Human Interleukin-7 in HIV-Infected Patients with low CD4 T cell Reconstitution on Antiretroviral Therapy: Results of two Phase II Multicentre Studies. 
\emph{Clinical Infectious Diseases} doi: 10.1093/cid/ciw065
\vspace{0.2cm}

[18] Guedj, J., Thi\'{e}baut, R. and Commenges, D. (2007b). 
Practical identiability of HIV dynamics models. \emph{Bulletin of mathematical biology} \textbf{69} 2493-2513.
\vspace{0.2cm}

[19] Genz, A. and Keister, B. (1996). Fully symmetric interpolatory rules for multiple integrals over innite regions with Gaussian weight. 
\emph{Journal of Computational and Applied Mathematics} \textbf{71} 299-309.
\vspace{0.2cm}

[20] Pinheiro, J.C. and Bates, D.M. (2000). \emph{Mixed-effects models in S and S-PLUS.} Springer.
\vspace{0.2cm}

[21] Drylewicz, J., Commenges, D. and Thiebaut, R. (2012). Maximum a Posteriori estimation in dynamical models of primary HIV infection. 
\emph{Statistical Communications in Infectious Diseases} \textbf{4}.
\vspace{0.2cm}

[22] Rue, H., Martino, S. and Chopin, N. (2009). Approximate Bayesian inference for latent Gaussian models by using integrated nested Laplace approximations. 
\emph{Journal of the royal statistical society: Series b (statistical methodology)} \textbf{71} 319-392.
\vspace{0.2cm}

[23] Van der Vaart, A.W. (2000). \emph{Asymptotic statistics} \textbf{3}. Cambridge university press.
\vspace{0.2cm}

[24] Prague, M., Commenges, D., Guedj, J., Drylewicz, J. and Thi\'{e}baut, R. (2013).
NIMROD: A program for inference via a normal approximation of the posterior in models with random effects based on ordinary differential equations. 
\emph{Computer methods and programs in biomedicine} \textbf{111} 447-458.
\vspace{0.2cm}

[25] Commenges, D., Jacqmin-Gadda, H., Proust, C. and Guedj, J. (2006).
A newton-like algorithm for likelihood maximization: The robust-variance scoring algorithm. \emph{arXiv preprint math/0610402}.
\vspace{0.2cm}

[26] Ramsay, J.O., Hooker, G., Campbell, D. and Cao, J. (2007). Parameter estimation for differential equations: a generalized smoothing approach. 
\emph{Journal of the Royal Statistical Society: Series B (Statistical Methodology)} \textbf{69} 741-796.
\vspace{0.2cm}

[27] Kuhn, E. and Lavielle, M. (2005). Maximum likelihood estimation in nonlinear mixed effects models. \emph{Computational Statistics $\&$ Data Analysis} \textbf{49} 1020-1038.
\vspace{0.2cm}

[28] Commenges, D., Joly, P., Anne, G.P. and Liquet, B. (2007). Choice between Semiparametric Estimators of Markov and Non-Markov Multi-state Models from 
Coarsened Observations. \emph{Scandinavian Journal of Statistics} \textbf{34} 33-52.
\vspace{0.2cm}

[29] Konishi, S. and Kitagawa, G. (2008). \emph{Information criteria and statistical modeling.}
Springer Science $\&$ Business Media.
\vspace{0.2cm}

[30] Commenges, D., Sayyareh, A., Letenneur, L., Guedj, J., Bar-Hen, A. et al. (2008).
Estimating a difference of Kullback-Leibler risks using a normalized difference of AIC.
\emph{The Annals of Applied Statistics} \textbf{2} 1123-1142.
\vspace{0.2cm}

[31] Commenges, D., Proust-Lima, C., Samieri, C. and Liquet, B. (2015). 
A universal approximate cross-validation criterion for regular risk functions. \emph{The international journal of biostatistics} \textbf{11} 51-67.
\vspace{0.2cm}

[32] Prague, M., Commenges, D., Drylewicz, J. and Thi\'{e}baut, R. (2012). 
Treatment Monitoring of HIV-Infected Patients based on Mechanistic Models. \emph{Biometrics} \textbf{68} 902-911.
\vspace{0.2cm}

[33] Mackall, C.L., Fry, T.J. and Gress, R.E. (2011). Harnessing the biology of IL-7 for therapeutic application. \emph{Nature Reviews Immunology} \textbf{11} 330-342.

\end{document}